# Spindle assembly checkpoint is sufficient for complete Cdc20 sequestering in mitotic control

Bashar Ibrahim

Bio System Analysis Group, Friedrich-Schiller-University Jena, and Jena Centre for Bioinformatics (JCB), 07743 Jena, Germany

Email: bashar.ibrahim@uni-jena.de

## Abstract

The spindle checkpoint assembly (SAC) ensures genome fidelity by temporarily delaying anaphase onset, until all chromosomes are properly attached to the mitotic spindle. The SAC delays mitotic progression by preventing activation of the ubiquitin ligase anaphase-promoting complex (APC/C) or cyclosome; whose activation by Cdc20 is required for sister-chromatid separation marking the transition into anaphase. The mitotic checkpoint complex (MCC), which contains Cdc20 as a subunit, binds stably to the APC/C. Compelling evidence by Izawa and Pines (Nature 2014; 10.1038/nature13911) indicates that the MCC can inhibit a second Cdc20 that has already bound and activated the APC/C. Whether or not MCC per se is sufficient to fully sequester Cdc20 and inhibit APC/C remains unclear. Here, a dynamic model for SAC regulation in which the MCC binds a second Cdc20 was constructed. This model is compared to the MCC, and the MCC-and-BubR1 (dual inhibition of APC) core model variants and subsequently validated with experimental data from the literature. By using ordinary nonlinear differential equations and spatial simulations, it is shown that the SAC works sufficiently to fully sequester Cdc20 and completely inhibit APC/C activity. This study highlights the principle that a systems biology approach is vital for molecular biology and could also be used for creating hypotheses to design future experiments.

## Introduction

Faithful DNA segregation, prior to cell division at mitosis, is vital for maintaining genomic integrity. Eukaryotic cells have evolved a conserved surveillance control mechanism for DNA segregation called the Spindle Assembly Checkpoint (SAC; [1]). The SAC monitors the existence of chromatids that are not yet attached correctly to the mitotic spindle and delays the onset of anaphase until all chromosomes have made amphitelic tight bipolar attachments to the mitotic spindle. A dysfunction in the SAC can lead to aneuploidy [2] and furthermore its reliable function is important for tumor suppression [3-4].

SAC acts by inhibiting the anaphase-promoting complex (APC/C or APC), a ubiquitin ligase, presumably through sequestering the ACP-activator Cdc20 (cf. Fig. 1A). APC activity is inhibited by the Mitotic Checkpoint Complex (MCC), which consists of the four checkpoint proteins Mad2, BubR1, Bub3, and Cdc20 [5]. A key MCC component is Mad2, a small protein that can adopt two conformations: 'open' inactive form (O-Mad2) and 'closed' active form (C-Mad2) [6-7]. C-Mad2 only forms when Mad2 binds to its kinetochore receptor, Mad1, or its checkpoint target Cdc20. The resulting C-Mad2-Cdc20 then binds to the BubR1–Bub3 complex, forming the MCC, which can then stably bind to the APC [5,8-9].

Furthermore, BubR1 has been suggested to interact with APC [10]. The complex Cdc20:C-Mad2 can also bind to the APC and form an inactive complex [11]. Another inhibitor, called the mitotic checkpoint factor 2 (MCF2), is associated with APC merely in the checkpoint arrested state but its composition is not known [12]. Recently and based on computational modeling, it has been shown that MCC alone is insufficient for fully inhibiting Cdc20 and APC. The same study has shown that cooperation between MCC and BubR1 is required to fully inhibit APC activity [13]. Very recent compelling evidence indicates that the MCC can inhibit a second Cdc20 that has already bound and activated the APC [14]. This data can enhance and elaborate on potential predictions from an integrative systems biology prospective.

So far, modelling of the SAC has helped to pinpoint advantages and problems of putative regulatory mechanisms [15-31]. These models can serve as a basis to integrate further findings and evaluate novel hypothesis related to checkpoint architecture and regulation. SAC models either consider few interacting elements using ordinary differential equations [18,21] or partial differential equations [15-17,29-30]; or conceive many interacting elements [20,22]. Other models use unconventional modelling approaches like Rule-Based modelling in space [26-27,29,32].

In this study a dynamical model for SAC activation and maintenance was constructed. This model considered all components of APC regulation in human cells in three variants: the MCC basic model variant, the MCC-BubR1 and the MCC that binds a second Cdc20 model variant. These models are validated with experimental data from the literature. A wide range of parameter values have been tested to find critical values of the APC binding rate. Simple mathematical analysis and computer simulations have helped to show that the MCC model variant in which MCC binds a second Cdc20 is sufficient to fully sequester Cdc20 and eventually completely inhibit APC activity.

## Materials and Methods

### Model assumptions

Some reactions can depend on the attachment status of the kinetochores, so all reactions can be classified by whether they are unaffected ("uncontrolled"), turned off ("off-controlled") or turned on ("on-controlled") upon microtubule attachment. Only reactions involving kinetochore localized species can be controlled. For example, formation of Mad1:C-Mad2:O-Mad2* (Reaction 2) can only take place as long as the kinetochores are unattached. In this model, if the kinetochore is unattached, u is set to u = 1, otherwise u = 0 [22-23]. Note that mass-action-kinetics is used for all reactions. Mad1:Mad2 is considered to be a preformed complex and the complex formation is not considered. It should be noted that this complex is a tetrameric 2:2 Mad1:Mad2 and not a monomer complex. From a mathematical point of view, considering the complex as a species would not make any difference in this case as long as there is one model. All previous mathematical models have considered the similar assumption to the template model (e.g., [18,20,23], see R1-R3).

For the spatial simulations, the mitotic cell is assumed as a 3D-ball with radius $R$. The last unattached kinetochore is a 2-sphere with radius $r$ in the center of the cell (Table 1). A lattice based model was used, which implies that the reaction volume of the mitotic cell is segmented into equal compartments. The initial concentrations of all freely diffuse species like Cdc20 and O-Mad2 are distributed randomly over all compartments of the mitotic cell. Localized species like Mad1:C-Mad2 and Mad1:C-Mad2:Mad2* are present at the kinetochore, their initial amount is located on the surface of the modeled 2-sphere. In order to observe a more accurate spatial behavior of the model variants, any symmetrical restrictions were not considered. All boundary conditions are reflective in order that the amount of particles is conserved.

**Numerical simulation of ODEs system**

The reaction rules are converted into sets of time dependent nonlinear ordinary differential equations (ODEs) by computing dS/dt = $\mathbf{N}\mathbf{v}$(S) with state vector S, flux vector v(S) and stoichiometric matrix $\mathbf{N}$. The actual initial amounts for reaction species are taken from literature (cf. Table 1). The kinetic rate constants ($k_{on}$ and $k_{off}$) are also taken from literature as far as they are known. In the other cases, representative values that exemplified a whole physiologically possible range were selected. A summary of all simulation parameters is given in Table 1. Also parameter scans were used to determine the critical and ideal rate values. In a typical simulation run, all reaction partners were initialized according to Table 1 and the ODEs were numerically solved until steady state was reached before attachment (using u = 1). After attachment, switching u to 0, the equations are again simulated, until steady state is reached.

The implementation and simulations code are written based on MATLAB (Mathworks, Natick, MA).

## Spatial simulation of PDEs system

Adding a second spatial-derivative as a diffusion term and a first-derivative as a convection term transforms the system of ODEs in coupled partial differential equations (PDEs) known as a r*eaction-diffusion-convection system* (see for details [29]).

Partial differential equations resulting from the reaction-diffusion-convection system were solved numerically using the open access Virtual Cell software [33]. The simulations are conducted using 3D geometries. Each dimension is divided into 51 parts, which results in 132.651 compartments in total. All parameters are set up consistent with the model assumptions. The system of PDEs with boundary and initial conditions is solved using the "Fully implicit finite volume with variable time-step" method. This method employs Sundials stiff solver CVODE for time stepping (method of lines) [33]. The derivations, necessary for diffusion and convection, are computed numerically. The human system is simulated for 1000 s which is sufficient to reach steady state, with a maximum time-step of 0.1 s and an absolute and relative tolerance of $1.0 \times 10^{-7}$. One simulation run takes between 1 and 10hs, dependent on the parameter-set. The time dependent concentration plots add up the amount of every species over all compartments and are generated with MATLAB (Mathworks, Natick, MA).

## Results

### Biochemical background of the model

The reaction network of the SAC activation and maintenance mechanism (Fig. 1) can be divided into three main parts: Mad2-activation template, MCC formation, and APC inhibition.

The essential component of the SAC-network is a kinetochore-bound template complex made up from Mad1 and C-Mad2. This template complex recruits O-Mad2 and stabilizes an intermediate conformation (O-Mad2*) which can bind Cdc20 efficiently and switches to closed conformation upon Cdc20-binding [6,34-35] (the biochemical equations of are described by reaction (R1-R3), Fig. 1, and reaction scheme). The C-Mad2-Cdc20 complexes formed by this mechanism, which has been given the name "template-model" [34], can further associate with the two proteins BubR1 (homologue of budding yeast Mad3) and Bub3 to form the tetrameric mitotic checkpoint complex (MCC; [5,36-37]). Another trimeric complex Bub3:BubR1:Cdc20 can form faster in the presence of unattached chromosomes [38] and it may be that MCC forms as an intermediate complex from which O-Mad2 rapidly dissociates [38-40]. The MCC and Bub3:BubR1:Cdc20 formations are described by the reaction (R4-R5, see chemical reaction scheme, below).

The APC is believed to be inhibited in multiple ways. Complexes of APC together with either Cdc20:C-Mad2 [11,41], Bub3:BubR1[10], Bub3:BubR1:Cdc20[10,41], MCC [5,8-9] or MCF2 [12] have been found to be inactive [12,38-40,42-43]. Recent work based on a systems biology approach, has shown that the MCC-BubR1 alone is able to reproduce both wild-type as well as mutation experiments of SAC mechanism. Hence these reactions, described by the reaction (R6-R7) (see chemical reaction scheme, below), are included. Free Cdc20 binds to and thereby activates the APC (R8) which promotes degradation of securin, which leads to cohesion cleavage by now active separase [44-46] (cf. Fig. 1B).

The following four model variants are considered: First is the core MCC model which consists of reactions (R1-R5, and R7, see also [20]). The second variant is the MCC-BubR1 model which consists of reactions (R1-R7, see also [13]). These model variants serve as the basic and reference models to compare with. The third and fourth model variants are the extension of the basic model variants with the addition of the MCC's ability to bind a second Cdc20 (R9, see also [14]).

**Chemical reaction scheme**

The SAC mechanism consists of 9 biochemical reaction equations describing the dynamics of the following 14 species: Mad1:C-Mad2, O-Mad2, Mad1:C-Mad2:O-Mad2*, Cdc20, Cdc20:C-Mad2, Bub3:BubR1, MCC, Bub3:BubR1:Cdc20, APC, MCC:APC, APC:BubR1:Bub3, APC:Cdc20:BubR1:Bub3, APC:Cdc20:MCC, and APC:Cdc20 (see also Fig. 1A).

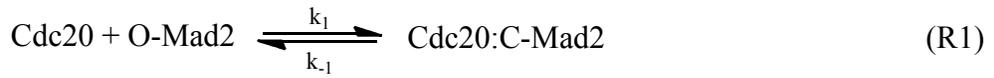

(R1)

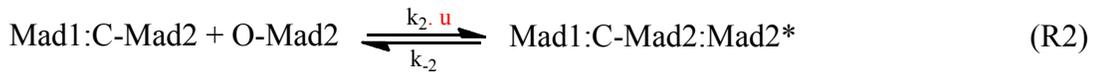

(R2)

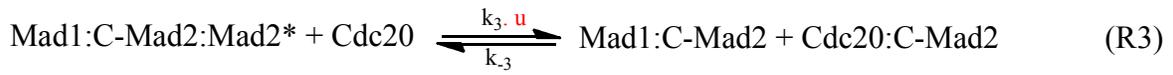

(R3)

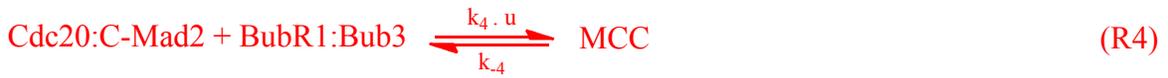

(R4)

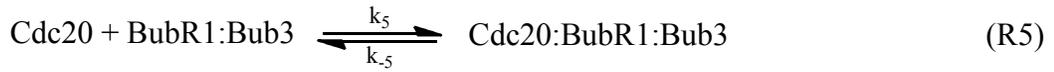

(R5)

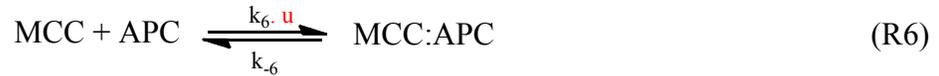

(R6)

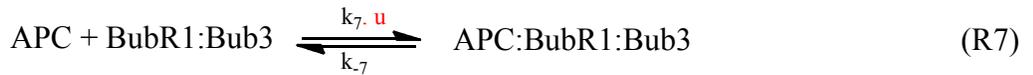

(R7)

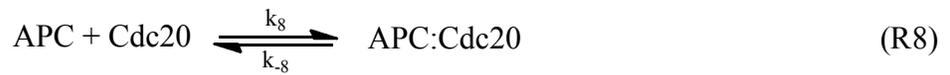

(R8)

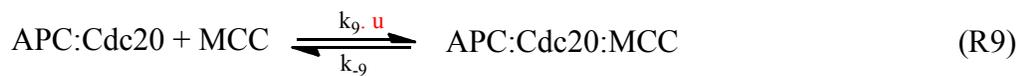

(R9)

**Time dependant dynamics of SAC regulation**

The SAC models, where either MCC is the exclusive inhibitor of APC [20] and both MCC together with BubR1, have been previously analyzed [13]. These models (see R1-R6, and R8; and R1-R8, respectively) are used in this study to build upon and to compare with the other two new model variants that mainly involve the ability of the MCC to bind a second Cdc20 that is already bound to APC [14]. In these two new model variants, in addition to the basic models variants (above), MCC binds a second Cdc20 that is already bound to APC (R9 is added).

Simulation results of these four model variants as non-linear ODEs are shown in Fig. 2. As for the Dynamics of free APC, all variants behave qualitatively similarly (Fig.1 left column). For the two new variants where MCC is able to bind a second Cdc20, slow MCC-APC binding rate is sufficient for fully APC:Cdc20 inhibition. Cdc20 sequestration reached about 95% with the MCC model variant that binds a second Cdc20 (Fig.1C middle column).

To validate all model variants, different mutations (deletion and over-expression) of the species involved were tested (Table 2). There are many experimental studies reported in the literature where deletion and also overexpression in different organisms of any of the core components, Mad2 [11,34,47-52], BubR1 [53-57], and Cdc20 [58-62], resulted in SAC failures, such as failed or successful mitotic arrest. These experiments may help in validating all model variants and additionally discriminating between them. The experiments from literature are listed in Table 2.

In the simulations, the respective initial concentration was set to zero for the deletions, and 100 fold higher concentrations for over-expression. The desired proper wild type functioning, APC:Cdc20 concentration should be very low (zero) before the attachment, and should increase quickly after attachment. Cells failing to arrest meant a very high level of APC:Cdc20 and low sequestration level of Cdc20. Arrested cells meant a very low level of APC:Cdc20 and full sequestration of Cdc20. The simulations show that all model variants are able to fully reproduce all known experimental findings for specific MCC-APC binding rate range (Fig. 3, Table 2). The simulations additionally indicate that the ideal MCC-APC binding rate for mutant type is $10^4$-$10^5$ $M^{-1}s^{-1}$ (Table 2).

Together, WT simulations of the four model variants were qualitatively similar. However, the variants where MCC binds a second Cdc20 did not require a high MCC-APC binding rate.

Additionally, the Cdc20 sequestration level is higher for the variant where MCC-binds a second Cdc20.

All model variants provided an ideal SAC functioning and were able to reproduce all experimental findings based on ODEs where only time but no space is included.

## Spatial dynamics of SAC regulation

Mathematical studies have recently shown that spatial properties such as diffusion and active transportation can play important roles in SAC activity and maintenance [17,23,29-30]. Lohel et al. [23] extended previously existing models of the SAC, to enable a detailed analysis of the kinetic consequences of localization. They found that the binding kinetics and stoichiometry are limiting factors for the overall dynamics of the SAC. Therefore, 3D space simulation was considered and spatial properties like diffusion and active transportation was tested with a range of kinetic reaction rates for APC binding. The environmental, diffusion and convection parameters are listed in Table1. For numerical simulation and geometry details see Materials and Methods.

The spatial simulation was run for each model variants four times. Three times as a reaction-diffusion system (Materials and Methods) for different MCC-APC binding rates; low rate $10^6$ M$^{-1}$s$^{-1}$ (Fig. 4, blue line), moderate rate $10^8$ M$^{-1}$s$^{-1}$ (Fig. 4, black lines) or high rate $10^{10}$ M$^{-1}$s$^{-1}$ (Fig. 4, red lines). Additionally, the simulation was run a fourth time to consider an active transportation for Mad2 as suggested by [29] as a reaction-diffusion-convection systems with a moderate MCC binding rate (see green dot lines in Fig. 4). Figure 4 depicts the wild type behavior of the average APC:Cdc20 concentration over time. All models should,in principle, be able to reproduce the desired behavior that is a very low level of APC:Cdc20. The MCC core model was able to reproduce the desired behavior only with a high MCC-APC binding rate or when convection is presented (Fig. 4A). These rates however are very high compared to the known SAC binding rate (e.g. Mad2-Cdc20 or Mad1-Mad2 [35,63]). The MCC-BubR1 core model variant was able to reproduce the desired behavior with any parameter set. However, the APC:Cdc20 was not fully inhibited (reached 90% of APC level, Fig. 4B). The new variants, in which MCC is able to bind a second Cdc20, were able to reproduce the desired behavior for any given parameter set and additionally were able to fully inhibit APC:Cdc20 activity (Fig. 4C-D). Only the MCC-BubR1 model variants that included additional Cdc20 binding, were able to reach steady state very fast (in about minute) while all other variants needed at least 3 minutes (Fig. 4A-D).

The level of free APC as well as free Cdc20 in each of the model variants was examined and these results are shown in Figure 5. The core models (MCC and MCC with BubR1) were able to sequester only 50% of Cdc20 amount (Fig.5A-B yellow lines). The new variants that included a second binding of Cdc20 were able to fully sequester Cdc20 together with the APC (Fig.5C-D, yellow and rose lines). However, the MCC-BubR1 variant that has additional Cdc20 binding was able to fully sequester both APC and Cdc20 after few seconds of the simulation (Fig.5D).

Taken together, the MCC-BubR1 model variant is able to capture ideal SAC behavior while not requiring very high binding rates or convection properties. Secondary Cdc20 binding [14] enhances SAC functioning.

## Discussion

Building on the investigation [20] of different models for Cdc20:Mad2 complex formation, the mathematical description of the SAC model have been enhanced by those reaction equations which describe additional Cdc20 sequestration by MCC, as reported recently [14]. A major role is played by the MCC and the BubR1, which in turn blocks APC activity. Four SAC model variants were analyzed; distinguishing the APC binding partners MCC or MCC and BubR1, and additionally the MCC's ability to bind a second Cdc20 that is already bound to APC. The latter succeeded to describe the correct metaphase to anaphase switching and also the ability to complete Cdc20 sequestering and APC inhibition. The calculations are in full agreement with the recent findings [14]. The model also indicated the value for the MCC-APC binding via a parameter scan (Fig. 6) and additionally favored the variant where both the MCC and BubR1 bind APC and additionally where the MCC binds a second Cdc20.

Computational modelling is a very important tool to elucidate how elaborate systems work. So far, mathematical models have helped to elucidate the kinetochore structure and with that the mitotic checkpoint mechanism [15-16,18-19,21,23-24,26-27,29,64]. These models mostly focus on either a minimal spatial model of SAC [15-16,23,29-30], namely the template model, or a detailed model excluding spatial effects [18-22]; consequently, previous models ignore the spatial and temporal regulation of multiple APC inhibition for SAC activity. In this work, both the approaches of using ODEs and PDEs were combined and this has enhanced the most detailed model available in the literature [22]. This work has also been confirmed by the very recent i experimental findings that the MCC binds a second Cdc20 [14].

In order to accelerate the pace of cell biology knowledge, systems analysis should be developed to link computational models of biological networks to experimental data in tight rounds of analysis and synthesis in an integrative systems biology framework. It is anticipated that such an approach for the SAC mechanism will serve as a basis to design experiments and evaluate novel hypotheses related to mitotic checkpoint control.

## Acknowledgment

The author would like to thank Fouzia Ahmad for proofreading the manuscript.This work was supported by the European Commission HIERATIC Grant 062098/14.

## Conflict of Interest

The author declares no conflict of interest.

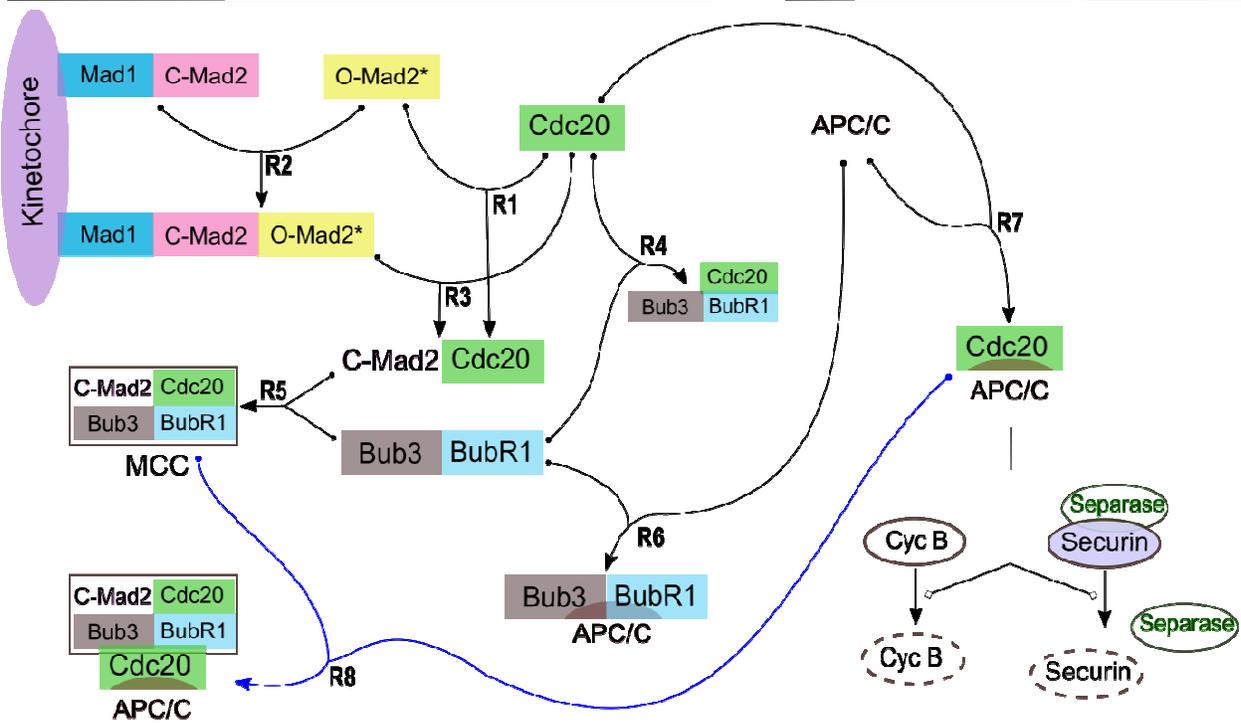

**Figure 1: Schematic representation of the core mechanism of SAC.**

(A) The SAC acts mainly through sequestration of the APC/C-activator Cdc20 by Mad2. Mad2 in closed conformation (C-Mad2) anchored at the kinetochore via Mad1 recruits cytosolic Mad2 in open conformation (O-Mad2). The so recruited Mad2 is stabilized in an intermediate conformation (Mad2*), which in turn is able to bind Cdc20 efficiently. The resulting C-Mad2-Cdc20 dimers are released from the kinetochore and form the mitotic checkpoint complex (MCC) together with Bub3 and BubR1. The Cdc20-containing complexes are not stable and dissociate with a certain rate, thus Cdc20 becomes available for APC/C activation soon after the last signaling Kinetochore is silenced by proper microtubule attachment. (B) When SAC signaling is turned off, Cdc20 binds to and thereby activates the APC/C. Active APC/C:Cdc20 promotes degradation of securin, which leads to cohesin cleavage by now active separase. The resulting separation of sister-chromatids is the hallmark of anaphase. Simultaneously, APC/C:Cdc20 promotes degradation of cyclin B, a requirement for mitotic exit.

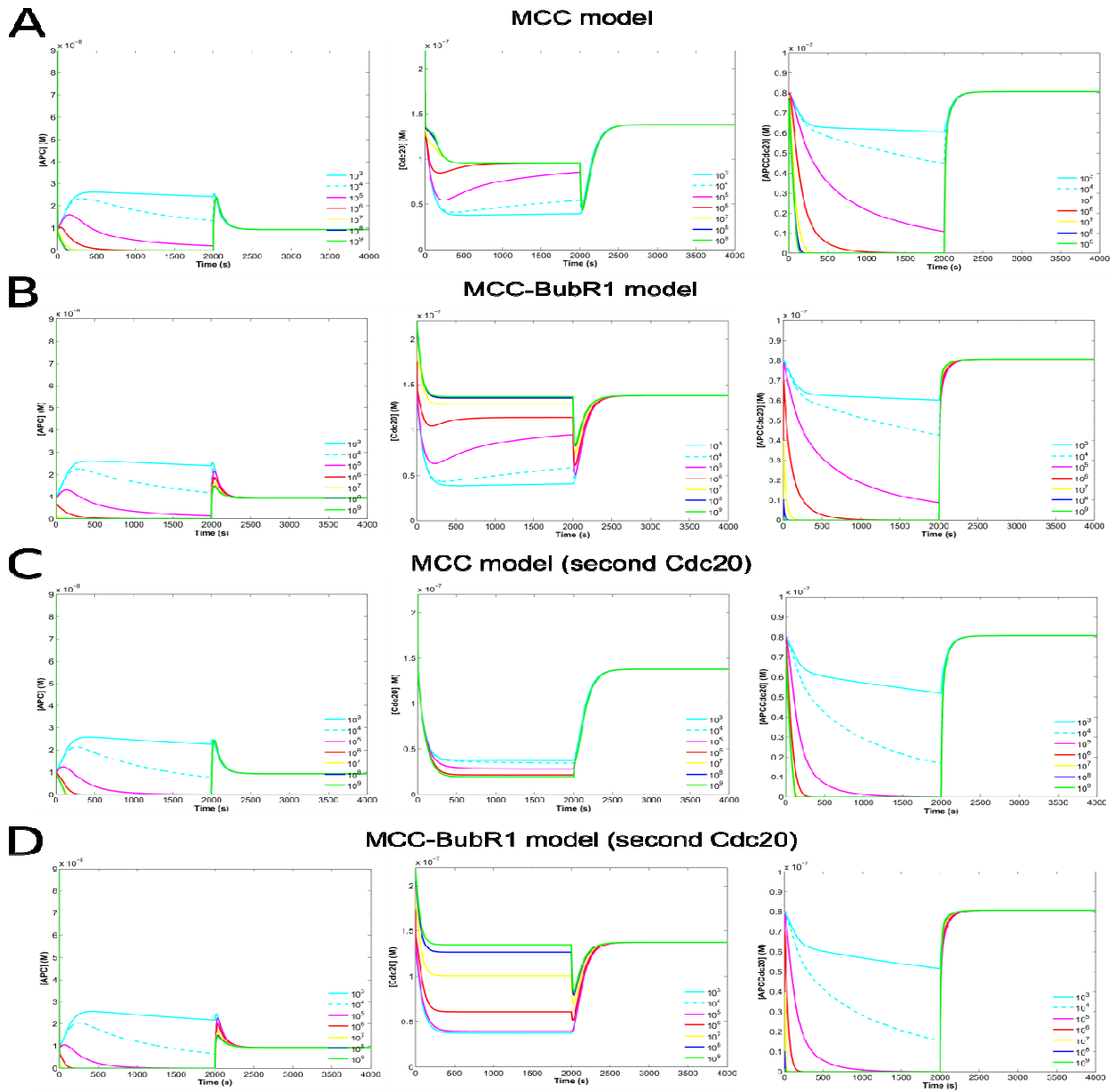

**Figure 2**: **Dynamical behavior of core SAC components concentration versus time**.
The columns from left to right show the APC, Cdc20, and APC:Cdc20 concentration (spindle attachment occurs at t = 2000s). All results are presented for different values of the rate $k_5$ (MCC binding to APC). Parameters setting are according to Table 1. Free APC concentration (left column) in all model variants is similar where its value at any given time is less than 30% of its initial concentration. The APC:Cdc20 dynamics (right column) in all model variants is also very similar, shows fast recovery and only with high MCC binding rate to APC shows fast inhibition for APC:Cdc20 activity. Cdc20 sequestration is depicted in the middle column. All model variants except the MCC-model variant that binds second Cdc20 (c.f. Panels A, B and D) are able to sequester about 80% of the free Cdc20 only with low MCC-APC binding rate. The MCC model variant that binds second Cdc20 (c.f. Panel C) is able to sequester around 95% of the free Cdc20 and independent of MCC-APC binding rate.

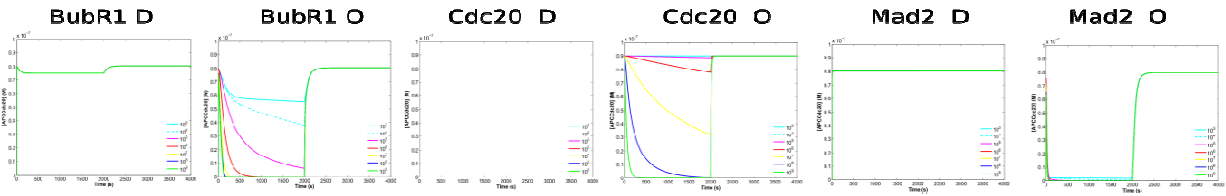

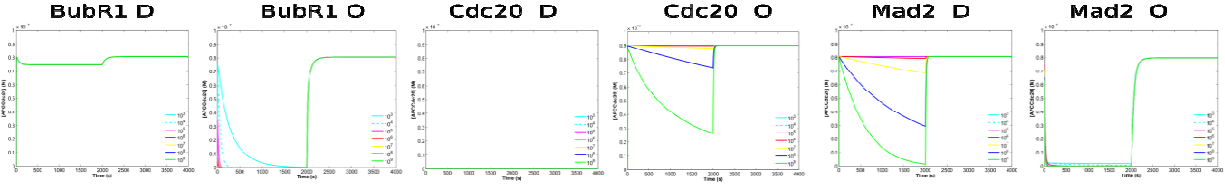

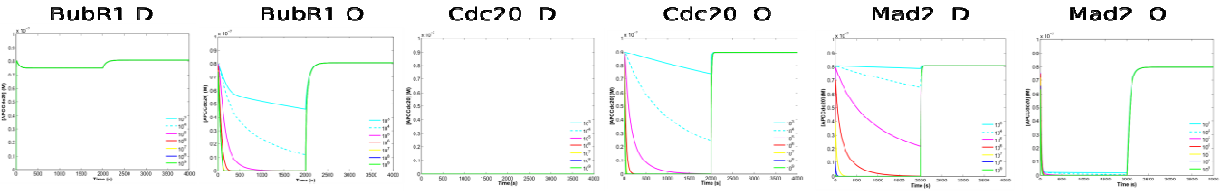

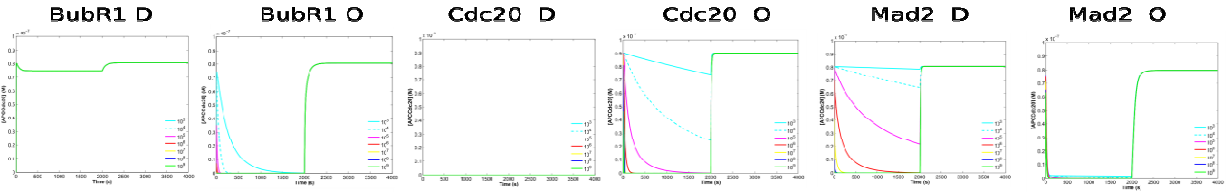

**Figure 3: Simulation of Mad2, BubR1, and Cdc20 mutations for each model variant.**
For deletion we set the respective initial concentration to zero, and for over-expression 100folds higher. Towards proper wild type functioning, APC:Cdc20 concentration should be very low (zero) before the attachment, and should increase quickly after attachment. Deletion of Mad2 or BubR1 or an overexpression of Cdc20 leads to inability of the cell to arrest, that is in the simulation, the concentration of APC:Cdc20 keeps high. Overexpression of Mad2 or BubR1 or deleting Cdc20 results in arresting the cell, that is, the concentration of APC:Cdc20 is very low or zero. Each row represents the mutation simulations of a model variant and a range of parameter rate for APC binding. Spindle attachment occurs at t = 2000s (switching parameter u from 1 to 0). All parameters setting are according to Table 1. See text for more details.

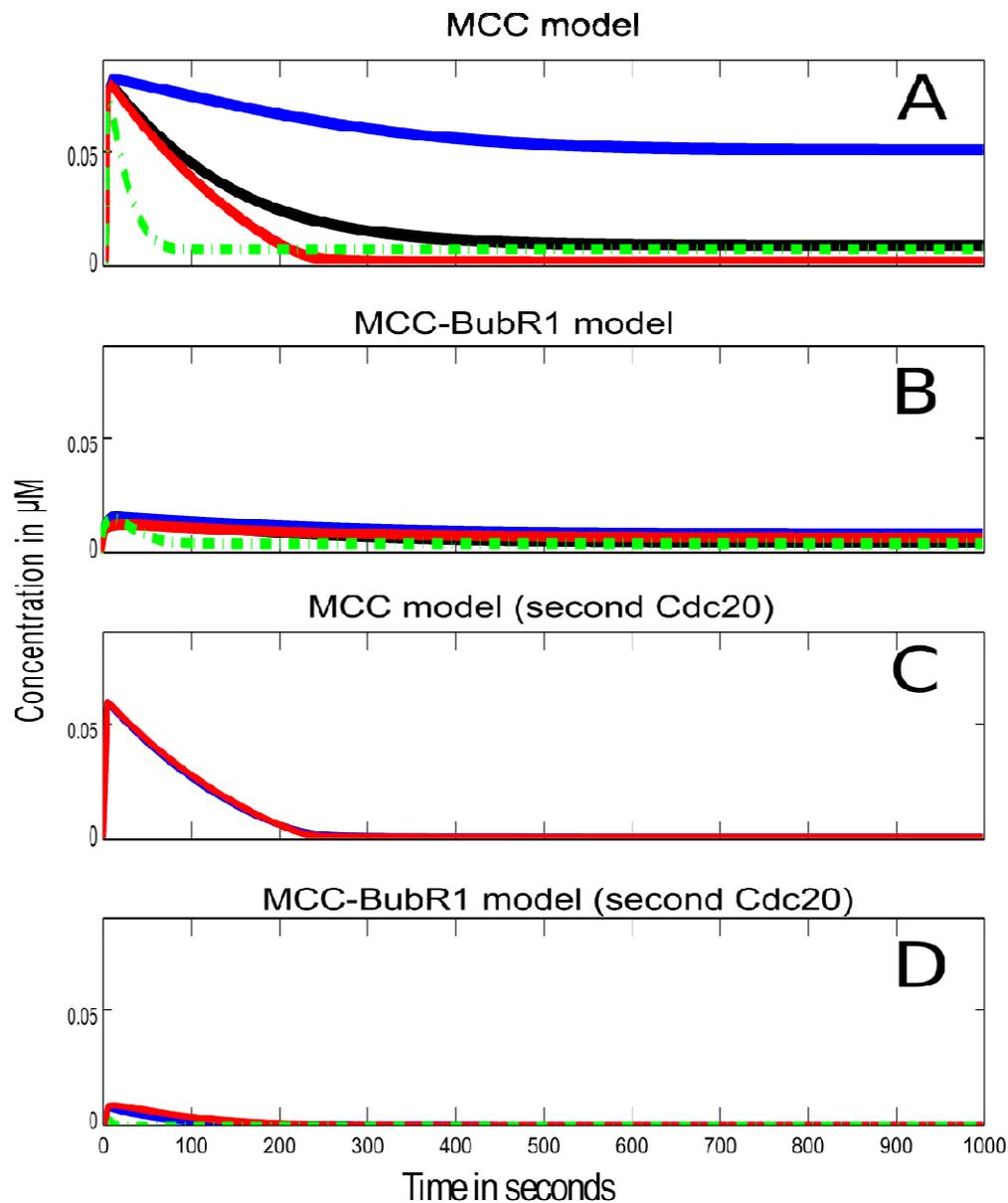

**Figure 4: Spatial simulation of SAC model variants.** The figures show the total concentrations over time for APC:Cdc20 with different parameter sets. All results are presented for different values of the APC/C binding rates ($k_5$, $k_6$ and $k_8$). Blue, black and red lines refer to the different APC/C binding rates, $10^6$ M$^{-1}$s$^{-1}$, $10^8$ M$^{-1}$s$^{-1}$ and $10^{10}$ M$^{-1}$s$^{-1}$ respectively. Dotted lines represent the simulations when Mad2 convection is included.

(A) Outcome of the simulated MCC core model (Reactions (1)-(5), and (7); cf. Table 1). It takes about 5 min to reach steady state except for the low rate value which takes 10 minutes. APC:Cdc20 is 90% inhibited only with high MCC-APC binding rate or when convection is included. (B) Outcome of the simulated MCC-BubR1 core model (Reactions (1)-(7); cf. Table 1). It takes about 3 min to reach steady state for any parameter set. (C) Outcome of the simulated MCC model that binds second Cdc20 (Reactions (1)-(5), (7),and (8); cf. Table 1). It takes about 3 min to reach steady state for any parameter set. (D) Outcome of the simulated MCC-BubR1 model that binds second Cdc20 (Reactions (1)-(8); cf. Table 1). It takes about 1.5 min to reach steady state for any parameter set.

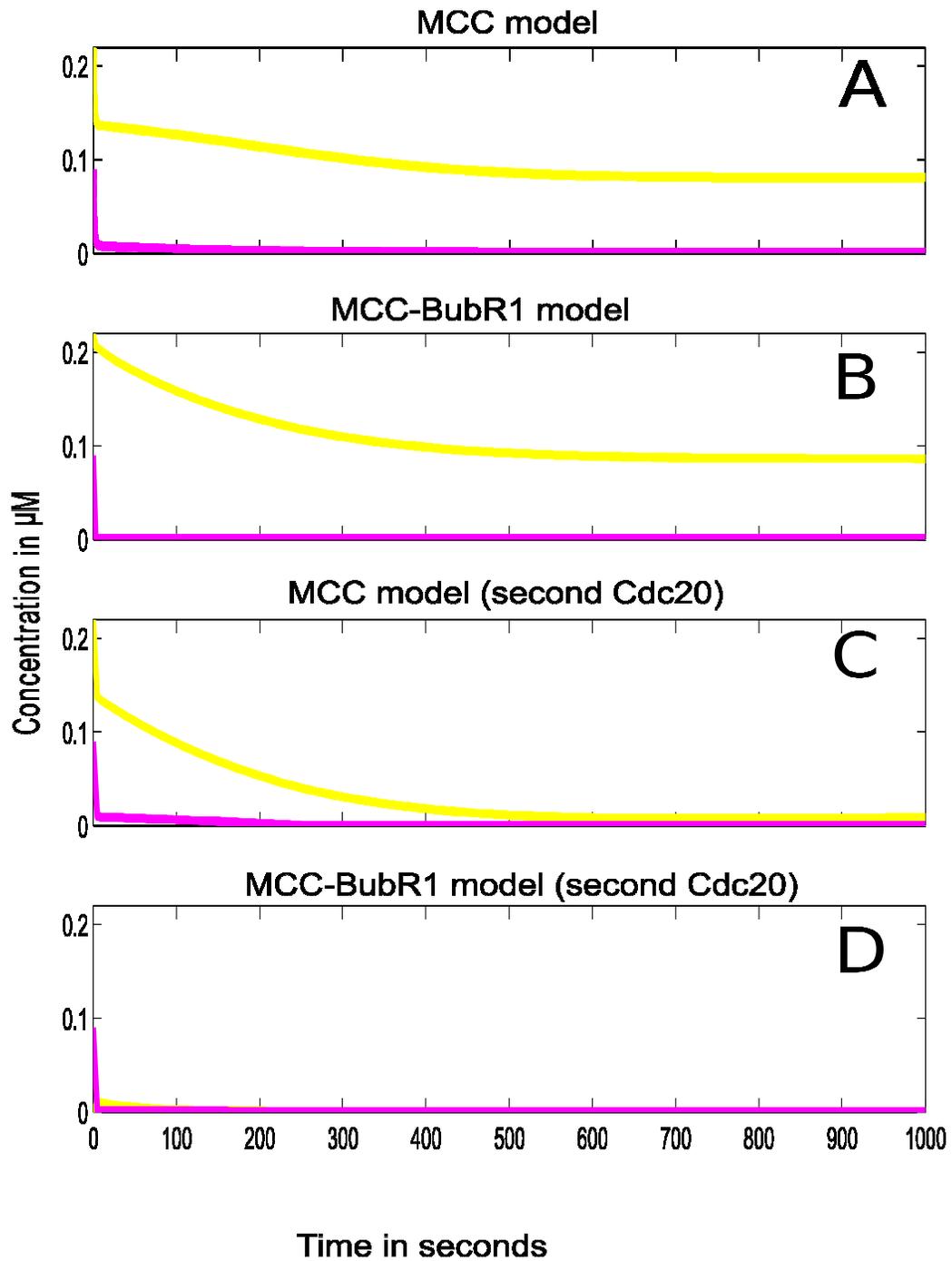

**Figure 5: Spatial simulation of APC and Cdc20 dynamics.** The figures show the total concentrations over time for free APC and free Cdc20. All results are presented for $10^8$ M$^{-1}$s$^{-1}$ value of the APC/C binding rates ($k_5$, $k_6$ and $k_8$). We can clearly see that only for the MCC-BubR1 model variant that binds second Cdc20, complete APC and Cdc20 sequestration is achieved.

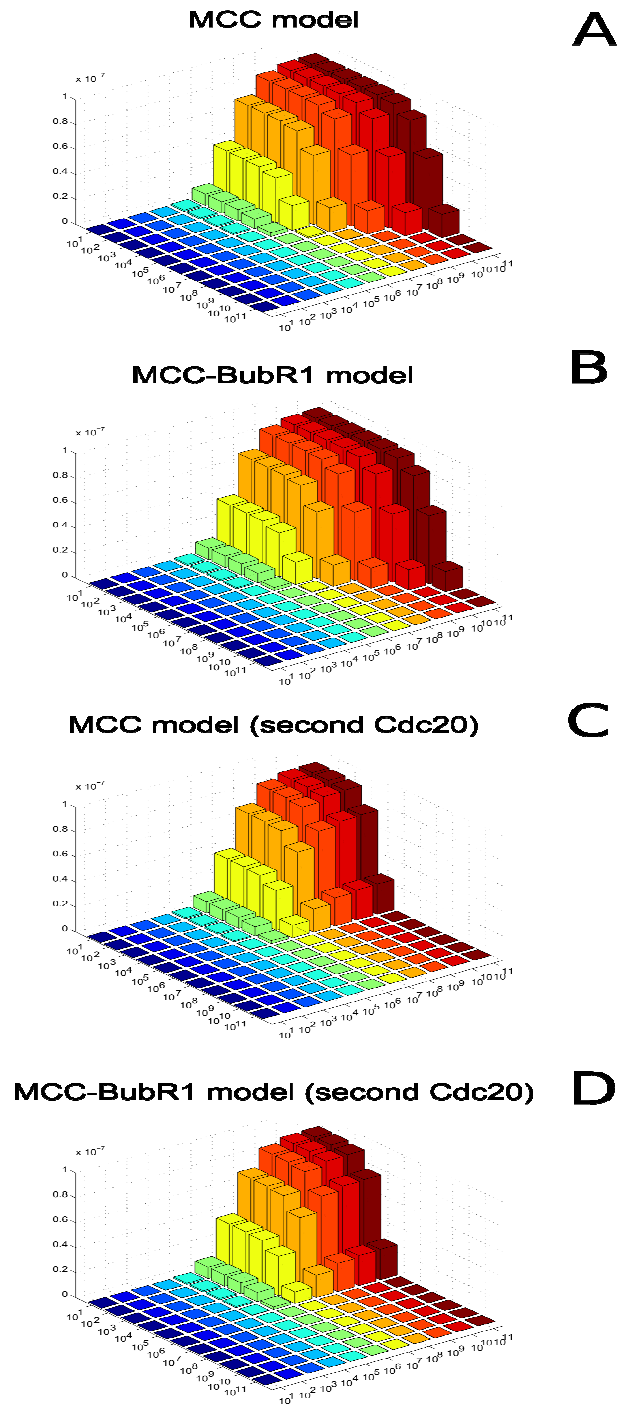

**Figure 6:** Sensitivities of the steady state concentrations of APC:Cdc20 associated with rate coefficients (k5 ) and (k7). Both parameters were varied in a range from $10^0$ M$^{-1}$s$^{-1}$ to $10^{11}$ M$^{-1}$s$^{-1}$. This analysis has been repeated for each model variants (panels, A, B, C, and D, respectively). Each model variant was simulated 121 times, and each simulation run until steady state reached. Panel A and B (core model variants) are very similar. The same is true for panel C and D (where MCC binds a second Cdc20). The scan of the new model variants, Panel C and D, indicate that the MCC-APC binding rate must be at least $10^5$ M$^{-1}$s$^{-1}$ and meanwhile APC-Cdc20 binding rate must not exceed $10^6$ M$^{-1}$s$^{-1}$.

**Table 1: Model parameters**

| | *Parameters* | | *Remarks* |
|---|---|---|---|
| **Rate constants** | | | |
| | $k_1$ | $1 \times 10^3$ $M^{-1}s^{-1}$ | [21,65] |
| | $k_2$ | $2 \times 10^5$ $M^{-1}s^{-1}$ | [35,63] |
| | $k_3$ | $1 \times 10^7$ $M^{-1}s^{-1}$ | [21] |
| | $k_4$ | $2 \times 10^4$ $M^{-1}s^{-1}$ | [21,29] |
| | $k_5$ | $10^3$-$10^9$ $M^{-1}s^{-1}$ | [20,29] |
| | $k_6$ | $10^3$-$10^9$ $M^{-1}s^{-1}$ | This study |
| | $k_7$ | $5 \times 10^6$ $M^{-1}s^{-1}$ | [20,22] |
| | $k_8$ | $10^3$-$10^9$ $M^{-1}s^{-1}$ | This study |
| | | | |
| | $k_{-1}$ | $1 \times 10^{-2}$ $s^{-1}$ | [21] |
| | $k_{-2}$ | $2 \times 10^{-1}$ $s^{-1}$ | [63] |
| | $k_{-3}$ | $0$ $s^{-1}$ | [21] |
| | $k_{-4}$ | $2 \times 10^{-2}$ $s^{-1}$ | [21,29] |
| | $k_{-5}$ | $1 \times 10^{-1}$ $s^{-1}$ | [29] |
| | $k_{-6}$ | $1 \times 10^{-2}$ $s^{-1}$ | This study |
| | $k_{-7}$ | $1 \times 10^{-1}$ $s^{-1}$ | [20,22] |
| | $k_{-8}$ | $8 \times 10^{-2}$ $s^{-1}$ | This study |
| **Initial amount** | | | |
| | Cdc20 | 0.22 $\mu$M | [41,66] |
| | O-Mad2 | 0.15 $\mu$M | [63] |
| | Mad1:C-Mad2 | 0.05 $\mu$M | [34] |
| | BubR1:Bub3 | 0.13 $\mu$M | [20,41,67] |
| | APC | 0.09 $\mu$M | [66] |
| | Other species start from zero | | |
| **Diffusion constants** | | | |
| | Cdc20 | 19.5 $\mu m^2 s^{-1}$ | [68] |
| | O-Mad2 | 5 $\mu m^2 s^{-1}$ | [29] |
| | Mad1:C-Mad2 | 0 $\mu m^2 s^{-1}$ | [29] |
| | Mad1:C-Mad2:Mad2* | 0 $\mu m^2 s^{-1}$ | [29] |
| | Bub3:BubR1 | 4 $\mu m^2 s^{-1}$ | [16,69] |
| | APC | 1.8 $\mu m^2 s^{-1}$ | [68] |
| | Other species diffusion coefficients are calculated from $D_{AB} = \frac{D_A * D_B}{D_A + D_B}$, where DA and DB are the diffusion coefficient for A and B, respectively. | | This study |
| **Convection constant** | | | |
| | O-Mad2 | 10 $\mu m s^{-1}$ | [29] |
| **Environment** | | | |
| | radius of the kinetochore | 0.1 $\mu$m | [70] |
| | radius of the cell | 10 $\mu$m | [29] |

**Table 2: *In-silico* mutation experiments for validation**

| Species | Exp. | Experimental effects | Effects in the model variants | | | |
|---------|------|----------------------|------------------|------------------|------------------|------------------|
| | | | **MCC core** | **MCC-BubR1 core** | **MCC extended** | **MCC-BubR1 extended** |
| BubR1 | D | SAC dysfunction [53-56] | Failed to arrest | Failed to arrest | Failed to arrest | Failed to arrest |
| BubR1 | O | Chromosomal instability [57] | Arrested $k_5 = 10^5$ | Arrested | Arrested $k_5 >= 10^4$ | Arrested |
| Mad2 | D | Cells are unable to arrest and impaired SAC (e.g., [11,47-50]) | Failed to arrest | Failed to arrest $k_5 = 10^6$ | Failed to arrest $k_5 = 10^4$ | Failed to arrest $k_5 = 10^4$ |
| Mad2 | O | Activates the SAC and blocks mitosis and stabilizes microtubule attachment [34,51-52] | Arrested | Arrested | Arrested | Arrested |
| Cdc20 | D | Cells arrested in metaphase [58-60] | Arrested $k_5 = 10^5$ | Arrested | Arrested | Arrested |
| Cdc20 | O | Impairment SAC and allows cells with a depolymerized spindle or damaged DNA to leave mitosis [61-62]. | Failed to arrest | Failed to arrest | Failed to arrest $k_5 =< 10^4$ | Failed to arrest $k_5 =< 10^4$ |

D refers to deletion or knockdown experiment, and O refers to an over-expression experiment. Failed to arrest means very high level of [APC:Cdc20] and low sequestration level of Cdc20. Arrested means very low level of [APC:Cdc20] and fully sequestration of Cdc20. Green means fully consistent with experiments and capture the desire behaviour. Yellow means consistent with experiments but required specific MCC-APC binding rate (see text for details)